\documentclass[showpacs,preprint,aps]{revtex4}
\usepackage{graphicx}
\usepackage{dcolumn}
\usepackage{bm}
\begin{document}
\setcounter{page}{1}
\title
{ Neutron structure effects in the deuteron and one neutron halos}
\author
{M. Nowakowski$^1$, N. G. Kelkar$^1$ and T. Mart$^2$}
\affiliation{$^1$ Departamento de Fisica, Universidad de los Andes,
Cra.1 No.18A-10, Santafe de Bogota, Colombia\\
$^2$ Departemen Fisika, FMIPA, Universitas Indonesia,
Depok 16424, Indonesia}
\begin{abstract}
Although the neutron ($n$) does not carry a total electric charge, its
charge and magnetization distributions represented in momentum space by the  
electromagnetic form factors, $F_1^{(n)} (q^2)$ and $F_2^{(n)} (q^2)$, 
lead to an electromagnetic potential of the neutron.  
Using this fact, we calculate the electromagnetic corrections to the 
binding energy, $B_d$, of the deuteron and a one neutron halo 
nucleus ($^{11}$Be), 
by evaluating the neutron-proton and the 
neutron-charged core ($^{10}$Be) potential,
respectively. The correction to $B_d$ ($\sim 9$ keV) is comparable to that
arising due to the inclusion of the $\Delta$-isobar component in the
deuteron wave function. In the case of the more loosely bound halo nucleus,
$^{11}$Be, the correction is close to about $2$ keV. 
\end{abstract}
\maketitle
\section{Introduction}
Studying the sub-structures of nuclei and nucleons, encoded either in
elastic electromagnetic form factors or deep inelastic structure functions, has been a 
field of continued interest in nuclear and particle physics. In this
paper we shall focus on the aspect of nucleon substructure in connection 
with the appearance and relevance of electromagnetic form factors in 
nuclear physics.
We shall calculate the electromagnetic contributions to the
binding energies of loosely bound neutron-nucleus systems, arising due to
the neutron form factors. Such systems are realized by the
deuteron and the one-neutron halo nuclei.    
Recently, the availability of radioactive beams opened up the 
possibility to study the structure of unstable nuclei. Such
experiments \cite{tani} revealed neutron rich nuclei whose spatial extensions 
are very large as compared to the range of the nuclear force. 
These so-called ``halo" nuclei consist of very loosely bound valence
neutrons which tunnel to distances
far from the remaining set of nucleons. 
Such nuclei can be viewed as a system of a ``core" with normal nuclear
density and a low density halo of one or more neutrons. 
Thus one expects that the neutrons in the halo do not experience the
strong force due to individual nucleons in the core but rather interact
with the core as a whole. Based on this understanding, many few body
models have been constructed and elaborately refined over the past
few years \cite{filo,others} in order to explain a variety of 
experimental data.

It is well known by now that the neutron
does have structure and hence an electric charge distribution
which can be measured in elastic electron-nucleon scattering experiments. 
The interest in this field was revived due to the new experimental
results on the nucleon form factor from the Jefferson laboratory 
\cite{jone}. 
Considering the recent interest in nucleon form factors along 
with the experimental advances being made in halo nuclear studies,   
we found it timely to investigate
the role of the neutron structure in the deuteron and one-neutron halo nuclei.
In what follows, we shall derive an expression for the electromagnetic
potential between the neutron and a charged particle (generalizing hereby
the Breit equation by adding to it the finite size corrections) 
and then apply it
to determine the perturbative corrections to the binding energy of 
nuclei.

\section{The neutron electromagnetic potential}
There exists a well known prescription for obtaining a potential from
quantum field theory \cite{lali4,fein}. The method consists of taking the 
Fourier transform of a non-relativistic scattering amplitude, say $M_{NR}$, 
for a scattering process of the type $A B \rightarrow A B$. 
Since the method is completely general, one can handle simple cases 
with an amplitude 
which corresponds to one or two particle exchange diagrams or 
consider more
complicated cases where the amplitude needs to be calculated using 
higher order corrections in the perturbative Feynman-Dyson expansion. 
If the amplitude depends only on the magnitude of momentum transfer ($Q$)
in the process, then the Fourier transform to obtain the potential, $V(r)$,  
reduces to \cite{marek1}:
\begin{equation}\label{fourtra}
V(r) = {1 \over (2\pi)^3} {1 \over r} 4\pi \, \int_0^{\infty}\, 
dQ \,Q\,\,M_{NR}(Q^2)\, \sin(Qr) \,.
\end{equation}
Obviously, if $M_{NR}(Q^2) \propto 1/Q^2$ (non-relativistic propagator of a 
massless particle), the potential is proportional to $1/r$. Some examples
of exotic potentials derived from 
quantum field theory can be found in \cite{fein,marek1,casi}.

The aim of the present work is to study the role of the neutron 
structure in neutron - charged particle interactions. Hence we shall 
be interested in obtaining an electromagnetic potential which 
describes the interaction between the charged particle and the 
neutron. To obtain this potential, we shall start with the scattering
amplitude for the process $n + A \rightarrow n + A$, where $A$ can be
any charged particle with charge $Ze$ (with $Z$ a positive integer). 
We shall consider two cases: the neutron (spin 1/2) + A (spin 1/2) and 
neutron (spin 1/2) + A (spin 0) case with the former being relevant for
the neutron-proton system (deuteron) and the latter for the one-neutron halo 
$^{11}$Be taken as a neutron plus a $^{10}$Be core in the $s$-state.

Mathematically, the complete spin-$1/2$ fermion vertex 
(fermion-photon-fermion) is given as \cite{marek3}:
\begin{eqnarray}
{\cal O}^{\mu} 
&=&F_1(q^2)\, \gamma^{\mu}\, +\, {i \sigma_{\mu \nu} \over 2m_f} \, 
q_{\nu}\, F_2(q^2) \, + i\epsilon^{\mu \nu \alpha \beta} {\sigma_{\alpha 
\beta} \over 4m_f} \, q_{\nu}\, F_3(q^2) \nonumber\\
&+& {1 \over 2m_f} \,\biggl(q^{\mu}\, - \, {q^2\over 2m_f} \gamma^{\mu} \biggr) \, 
\gamma_5 \, F_4(q^2)~,
\end{eqnarray}
where the sub-structure of the fermion is contained in the various
form factors, $F_i$ ($i=1,2,3,4$). $q_\mu$ is the four-momentum carried by the
photon ($q^2=q^{\mu}q_{\mu}$), $m_f$ is the mass of the fermion 
and the $\gamma$'s are the usual 
Dirac matrices \cite{bjor}. 
The four-momentum squared, $q^2 = \omega^2 - \vec{Q}^2$ 
(where $\omega$ is the energy and $\vec{Q}$ the three momentum), reduces
in the non-relativistic limit ($\omega \sim 0$) to $q^2 = -\vec{Q}^2$. 
The above four form factors appear in the expression for the cross section
for electron-nucleon elastic scattering and the nucleon form 
factors are thus extracted 
from such scattering experiments. The first form factor, $F_1$ is connected 
to the charge distribution inside the nucleon and $F_1(0) =$ the total
charge of the nucleon. The form factors $F_2$, $F_3$ and $F_4$ are 
related to the anomalous magnetic moment, electric dipole and Zeldovich 
anapole moment, respectively. 
In what follows, we shall obtain the electromagnetic potential for the 
spin-1/2 - spin-1/2 and the spin-1/2 - spin-0 case in terms of the 
nucleon electromagnetic form factors $F_1(q^2)$ and $F_2(q^2)$ as the 
other form factors ($F_3(q^2)$ and $F_4(q^2)$) are related to the 
weak interaction and their effects are small. 

\subsection{The neutron proton case}
\label{subsec:np_case}
\begin{figure}[h]
\includegraphics[width=12cm,height=8cm]{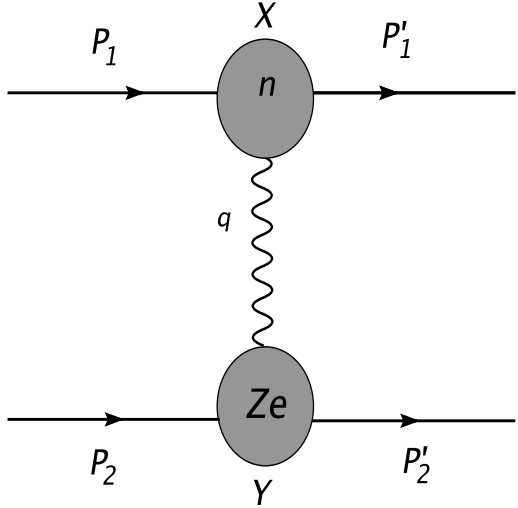}
\caption{\small Feynman diagram for neutron scattering from a nucleus with 
$Z$ being the number of protons.}
\end{figure}
In order to perform a complete calculation of the potential, we turn
to the task of expanding the amplitude in $1/c^2$ terms, thereby
generalizing the Breit equation \cite{lali4} by the inclusion
of two electromagnetic form factors. This will lead to non-local
and spin-dependent terms in the potential whose 
contributions to the binding energies are not negligible as will be seen later. 
Starting with the Feynman diagram of Fig. 1, the 
nucleon-photon-nucleon vertices in terms of the nucleon form factors 
$F_1$ and $F_2$ can be written as,
\begin{eqnarray}\label{vertices}
\Gamma_X^{\mu} &= &F_1^X\, \gamma^{\mu}\,\, - \, \, {\sigma^{\mu \nu} 
\over 2 M_X\, c}\,q_{\nu}\, F_2^X \\ \nonumber
\Gamma_Y^{\mu} &= &F_1^Y\, \gamma^{\mu}\,\, + \, \, {\sigma^{\mu \nu} 
\over 2 M_Y\, c}\,q_{\nu}\, F_2^Y \,.
\end{eqnarray}
The photon momentum, $q = P_1^{\prime} - P_1 = P_2 - P_2^{\prime}$. In 
the non-relativistic limit ($q^0 = 0$), $q^2 = - \vec{Q}^2$, where, 
$\vec{Q} = \vec{p}_1^{\,\,\prime} - \vec{p}_1 = \vec{p}_2 - \vec{p}_2^{\,\, \prime}$.
The amplitude for the process $ N + N \rightarrow N + N$ is then given by
\cite{lali4},
\begin{equation}
M_{fi} = e_X \,e_Y\, \biggl [ \, \biggl ( \bar{u}(\vec{p}_1^{\,\,\prime})\, 
\Gamma_X^{\mu}\, u(\vec{p}_1) \biggr )\, \, D_{\mu \nu}(\vec{Q})\,\, \biggl ( 
\bar{u}(\vec{p}_2^{\,\,\prime})\, \Gamma_Y^{\nu}\, u(\vec{p}_2) \biggr )\,
\biggr]\,
\, ,
\end{equation}
where, 
$D_{\mu \nu}(\vec{Q})$ is the photon propagator and 
$u(\vec{p}_1)$, $u(\vec{p}_2)$ etc., the usual Dirac spinors given as,
\begin{equation}\label{4}
u \,=\,\sqrt{2M}  \left[
\begin{array}{c}
{\displaystyle \biggl ( 1 - {\vec{p}^{\,2} \over 8 M^2c^2} \biggr ) \,w}
\\ [0.5em]
{\displaystyle {\vec{\sigma} \cdot \vec{p} \over 2 M c} \, w}
\end{array}
\right ]\,.
\end{equation}
Substituting for the $u(\vec{p})$'s and the vertex factors given above, the 
amplitude is evaluated and then rearranged to be written in the following form:
\begin{equation}
M_{fi} = - 2 \,M_X\,\cdot \,2\,M_Y\,[w_1^{\prime *}\,w_2^{\prime *}\,\,
U(\vec{p}_1, \vec{p}_2, \vec{Q})\,w_1\,w_2\,]\,,
\end{equation}
thus obtaining the potential $U(\vec{p}_1, \vec{p}_2, \vec{Q})$ in momentum 
space. The potential in $r$-space is obtained simply from a Fourier transform
of the potential in momentum space, namely,
\begin{equation}
V(\vec{r}) \,=\, \int \,e^{i\vec{Q}\cdot\vec{r}}\,\,U(\vec{p}_1, \vec{p}_2, \vec{Q}) 
\,\,{d^3q \over (2\,\pi)^3}\,\,.
\end{equation}
The electromagnetic neutron-proton potential in momentum space, using the
above procedure is found to be,
\begin{eqnarray}\label{fullpotnp}
&&U^{np}(\vec{p}_1, \vec{p}_2, \vec{Q}) = 4 \pi e^2 \,\biggl \{ \,F_1^n\,F_1^p\,
\biggl [\,{1 \over \vec{Q}^2} \,-\,{1 \over 8 \,M_n^2\,c^2}\, -\, 
{1 \over 8 \,M_p^2\,c^2}\,
-\, {\vec{\sigma}_1 \cdot \vec{\sigma}_2
\over 4 \,M_n\,M_p\,c^2}\nonumber\\ \nonumber
&&+\,\, {(\vec{\sigma}_1 \cdot \vec{Q}) 
(\vec{\sigma}_2 \cdot \vec{Q}) \over 4 \,M_n\,M_p\,c^2 \vec{Q}^2} 
 +\, {i \vec{\sigma}_1 \cdot (\vec{Q} \times \vec{p}_1) \over 4\,M_n^2\,c^2\,
\vec{Q}^2}\,-\, {i \vec{\sigma}_2 \cdot (\vec{Q} \times \vec{p}_2) \over 4\,M_p^2\,
c^2\,\vec{Q}^2}\,+\, {i \vec{\sigma}_2 \cdot (\vec{Q} \times \vec{p}_1) \over 
2\,M_p\,M_n\,c^2\,\vec{Q}^2}\,
-\, {i \vec{\sigma}_1 \cdot (\vec{Q} \times \vec{p}_2) \over 
2\,M_p\,M_n\,c^2\,\vec{Q}^2} \\ \nonumber
&&-\, {\vec{p}_1\cdot \vec{p}_2 \over M_n\, M_p\, 
c^2\,\vec{Q}^2} \,
+\,{(\vec{p}_1 \cdot \vec{Q})\,(\vec{p}_2 \cdot \vec{Q}) 
\over M_n\,M_p\,c^2\,\vec{Q}^4}\,\biggr ] \\ \nonumber
&&+\,F_1^n\,F_2^p\,\biggl [\,-\,
{1 \over 4\,M_p^2\,c^2} \, -\, 
{\vec{\sigma}_1 \cdot \vec{\sigma}_2
\over 4 \,M_n\,M_p\,c^2}\,+\,\, {(\vec{\sigma}_1 \cdot \vec{Q}) 
(\vec{\sigma}_2 \cdot \vec{Q}) \over 4 \,M_n\,M_p\,c^2 \vec{Q}^2}\, 
+\, {i \vec{\sigma}_2 \cdot (\vec{Q} \times \vec{p}_1) \over 
2\,M_p\,M_n\,c^2\,\vec{Q}^2}\,-\,{i \vec{\sigma}_2 \cdot
(\vec{Q} \times \vec{p}_2) \over 2\,M_p^2\,
c^2\, \vec{Q}^2}\,\biggr] \\ \nonumber 
&&+\,F_2^n\,F_1^p\,\biggl [\,-\,
{1 \over 4\,M_n^2\,c^2} \, -\, 
{\vec{\sigma}_1 \cdot \vec{\sigma}_2
\over 4 \,M_n\,M_p\,c^2}\,+\,\, {(\vec{\sigma}_1 \cdot \vec{Q}) 
(\vec{\sigma}_2 \cdot \vec{Q}) \over 4 \,M_n\,M_p\,c^2 \vec{Q}^2}\, 
-\, {i \vec{\sigma}_1 \cdot (\vec{Q} \times \vec{p}_2) \over 
2\,M_p\,M_n\,c^2\,\vec{Q}^2}\,+\,{i \vec{\sigma}_1 \cdot 
(\vec{Q} \times \vec{p}_1) \over 2\,M_n^2\,
c^2\, \vec{Q}^2}\,\biggr] \\ 
&&+\,F_2^n\,F_2^p\,\biggl [ \,-\, {\vec{\sigma}_1 \cdot \vec{\sigma}_2
\over 4 \,M_n\,M_p\,c^2}\,+\,\, {(\vec{\sigma}_1 \cdot \vec{Q}) 
(\vec{\sigma}_2 \cdot \vec{Q}) \over 4 \,M_n\,M_p\,c^2 \vec{Q}^2}
\,\biggr ]\,\,\biggr \}\,.
\end{eqnarray}
In obtaining the above expression, we have dropped terms of order 
higher than ($1/c^2$). The above expression involves a separate 
$1/c^2$ expansion corresponding 
to each of the form factor combinations. 
This is to say, the term proportional to
$F_1^n F_1^p/Q^2$ is the leading term of the coupling constants combination
$F_1^n F_1^p$, whereas the other terms proportional to this combination
are $1/c^2$ suppressed. However, we cannot say that just like 
the ($1/c^2$) terms in $F_1^n F_1^p$, those in the 
other combinations like $F_1^n F_2^p$, $F_2^n F_1^p$ and 
$F_2^n F_2^p$ are also suppressed, as the latter 
represent different combinations of coupling constants. Indeed, all 
combinations different from $F_1^n F_1^p$ are displayed also in the leading 
order albeit they appear with the 
$1/c^2$ factor which does not imply that they are smaller
than $F_1^n F_1^p/Q^2$. We will see later how this works in practice.
The constant $e$ in the above is
defined such that $e^2 = \alpha$, the fine structure constant.
Since the experimental form factors for the neutron as well as the proton 
are normalized in a similar way, we have a vertex factor of
$\sqrt{4 \pi \alpha}$ at the neutron-photon-neutron vertex too, in 
order to remove the extra factor by which the experimental 
neutron form factor was divided.

The detailed calculation of the Fourier transforms of the terms in 
(\ref{fullpotnp}) and the corresponding potential in $r$-space is 
given in the appendix. The terms which appear with cross 
products of momenta in (\ref{fullpotnp}), 
do not contribute to the binding energy correction for the deuteron, 
as shown in the appendix.

\subsection{The neutron spin-zero-nucleus case}
To evaluate the electromagnetic potential between the neutron and a 
spin-zero nucleus, we repeat a very similar procedure as above, with the
vertex $Y$ in Fig. 1 replaced in this case by,
\begin{equation}
\Gamma_Y^{\nu} = F^Y(q^2) \,(P_2 + P_2^{\prime})^{\nu}\, ,
\end{equation}
where, $F^Y(q^2)$ is the form factor of the nucleus ($A$) in momentum space.
We also replace the appropriate normalization for the spin zero particles.
The potential in momentum space is found to be,
\begin{eqnarray}\label{potfullnuclear}
U^{nA}(\vec{p}_1, \vec{p}_2, \vec{Q}) = 4 \pi e^2 \,\biggl \{ \,F_1^n\,F^Y\,
\biggl [\,{1 \over \vec{Q}^2} \,-\,{1 \over 8 \,M_n^2\,c^2}\, 
-\, {\vec{p}_1\cdot \vec{p}_2 \over M_n\, M_A\, 
c^2\,\vec{Q}^2}
+\,{(\vec{p}_1 \cdot \vec{Q})\,(\vec{p}_2 \cdot \vec{Q}) 
\over M_n\,M_A\,c^2\,\vec{Q}^4} \nonumber \\ \nonumber
 +\, {i \vec{\sigma}_1 \cdot (\vec{Q} \times \vec{p}_1) \over 4\,M_n^2\,c^2\,
\vec{Q}^2}\,
-\, {i \vec{\sigma}_1 \cdot (\vec{Q} \times \vec{p}_2) \over 
2\,M_A\,M_n\,c^2\,\vec{Q}^2} \biggr ] \\ 
+\,F_2^n\,F^Y\,\biggl [ -\,{1 \over 4 \,M_n^2\,c^2}\,
 +\, {i \vec{\sigma}_1 \cdot (\vec{Q} \times \vec{p}_1) \over 2\,M_n^2\,c^2\,
\vec{Q}^2}\,
-\, {i \vec{\sigma}_1 \cdot (\vec{Q} \times \vec{p}_2) \over 
2\,M_A\,M_n\,c^2\,\vec{Q}^2} \biggr ] \,\biggr\}\, .
\end{eqnarray}
This potential will be used in the present work to evaluate the
correction to the binding energy of the one-neutron halo nucleus,
$^{11}$Be, taken as a neutron plus $^{10}$Be core. The evaluation of
the $^{11}$Be wave function is usually done in a neutron plus $^{10}$Be 
core model where the ground state (spin zero) and an excited 
state (spin 2) of the core are considered. We shall restrict to assuming 
the core to be a spin zero nucleus since the calculation of a 
spin 1/2 - spin 2 electromagnetic potential is beyond the scope of the
present work. The contributions of the spin-2 terms to the $^{11}$Be 
wave function are in any case small and would have only a small overlap
with the electromagnetic potential which is not long-ranged.

In the next sub-section, we shall describe the various parameterizations of
the form factors which will be used to evaluate the above potentials.

\subsection{The electromagnetic form factors $F_1(q^2)$ and $F_2(q^2)$}
The form factors $F_1$ and $F_2$ have been extracted from several 
experiments and parameterized in different forms in the past \cite{gasi}. 
However, a discrepancy between the old results and new 
experiments which extract the two form factors through polarization 
measurements was recently reported \cite{jone}. It
was found, however, that this discrepancy could be resolved by taking into
account the two photon contribution \cite{chen}. Therefore, 
in the following, we shall make use of the `standard' parameterization of the
form factors. Considering the 
renewed debate on the nucleon form factors, in the present work
we shall also perform the calculations with different parameterizations
available in the literature. The Sachs form factors $G_E$ and $G_M$ which 
appear in the expressions for the electron-nucleon elastic cross sections 
are related to the structure functions $F_1$ and $F_2$. The nucleon 
form factors $F_1^N(q^2)$ and $F_2^N(q^2)$ are thus given as  
\begin{eqnarray}\label{formf1}
F_1^N(q^2) = {4 M^2 \, G_E^N(q^2) \, - \, q^2\, G_M^N(q^2) 
\over 4M^2 \,-\, q^2} ~,\\ \nonumber
F_2^N(q^2) = {4 M^2 [\, G_M^N(q^2) \, - \, G_E^N(q^2)] 
\over 4M^2 \,-\, q^2} 
\end{eqnarray}
where $M$ is the mass of the nucleon and $q^2 = \omega^2 - \vec{Q}^2$ 
is the four momentum of the virtual photon as mentioned before. 
A large body of experiments starting from the sixties until now has 
been dedicated to the extraction of $G_E^n(q^2)$ and $G_M^n(q^2)$. Though 
the newer experiments have smaller statistical errors, there still exist
uncertainties arising from the theoretical description of the deuteron. 
Hence, in the case of the neutron, 
we shall use different parameterizations available in literature.

\subsubsection{The dipole form factor and the neutron electric potential}
We shall
use the following form for $G_E^n(q^2)$ \cite{plat} with different
sets of parameters `$a$' and `$b$' obtained in literature:
\begin{equation}\label{bostedge}
G_E^n(q^2) = {a \,\mu_n \, (q^2/4M^2) \over 1 \, - \, b\, (q^2/4M^2)} 
\, G_D (q^2) ~,
\end{equation}
where $\mu_n$ is the neutron magnetic moment and $G_D (q^2)$ is the
standard ``dipole fit" which is generally used in summarizing the 
electron-nucleon elastic scattering data. With $G_D (q^2)$ defined as
\begin{equation}
G_D (q^2) = {1 \over (1 \,-\, q^2/m^2)^2 } ~,
\end{equation}
it was observed that 
\begin{equation}
G_E^p(q^2) \simeq {G_M^p(q^2) \over \mu_p} 
\simeq {G_M^n (q^2)\over \mu_n} \simeq G_D(q^2) ~,
\end{equation}
where the magnetic moments $\mu_p$ and $\mu_n$ of the proton and neutron
(in nuclear magneton) 
are 2.79 and $-1.91$, respectively. 
Using the above definitions of $G_E^n(q^2)$ and $G_M^n(q^2)$ in 
(\ref{formf1}), we obtain 
the following non-relativistic expression (i.e. with $q^2 \simeq -\vec{Q}^2$) 
for the form factor $F_1^n(Q^2)$:
\begin{equation}\label{nonrelff}
F_1^n(Q^2) = |\mu_n|\, Q^2 \left [ {4M^2\,(a-1)\, - \,Q^2b \over 
4 M^2\, + \,bQ^2} \right ] \,\,{1 \over 4M^2 + Q^2} \, \, 
{1\over (1 + Q^2/m^2)^2}\,.
\end{equation}

At this point, it is nice to note that using the dipole form factors,
the potential between the 
structured neutron and an electron can be derived
analytically and is given by, 
\begin{equation}\label{pot}
V^n(r) = {- Z \alpha \,|\mu_n| \over (1-b)}\,{1 \over r}\,\,
\biggl[ \,(a-1)\,H_1^n \,-b\,H_2^n\,\biggr ] ~,
\end{equation}
where
\begin{eqnarray}
&&H_1^n = {1 \over (\kappa^2 - 1)^2} \,e^{-\kappa m r} \,-\, 
{1 \over (\kappa^{\prime 2} - 1)^2} \,e^{-\kappa^{\prime} m r}\nonumber\\ 
&+& \left[{1 \over (\kappa^{\prime 2} - 1)^2} \, -\, 
{1 \over (\kappa^2 - 1)^2} \right]\,e^{-mr}\,+\, {1\over 2}\,\left[ 
{m \over (\kappa^2 - 1)}\,-\, {m \over (\kappa^{\prime 2} - 1)}\right]
\,r\,e^{-mr} 
\end{eqnarray}
and 
\begin{eqnarray}
H_2^n &=& -{1 \over (\kappa^2 - 1)^2}\,e^{-\kappa m r}\,+\, 
{1 \over (\kappa^{\prime 2} - 1)^2}\,\biggl({\kappa^{\prime}\over \kappa}
\biggr)^2\,e^{-\kappa^{\prime} m r} \nonumber\\ \nonumber
&-& \,\left[ {1 \over (\kappa^{\prime 2} - 1)^2} \,-\, 
{1 \over (\kappa^2 - 1)^2} \right] \,{1 \over \kappa^2} \,e^{-mr}\,
\\ 
&-& \,\left[ {1 \over (\kappa^2 - 1)}\,
-\, {1 \over (\kappa^{\prime 2} - 1)} \right] \,{1 \over \kappa^2} \,e^{-mr} \biggl(
{mr \over 2} - 1\biggr) ~,
\end{eqnarray}
where $\kappa = 2M/m$, $\kappa^{\prime} = \kappa/\sqrt b$ 
and typically $m^2 =0.71$ GeV$^2$ \cite{gasi,bosted}.

In \cite{bosted}, the neutron magnetic form factor was also fitted 
with the form 
\begin{equation}\label{bostgm}
{G_M^n(Q^2) \over \mu_n} = {1 \over 1 \,-\,1.74Q\,+\,9.29Q^2\,-\,7.63Q^3\,
+\,4.63Q^4} ~,
\end{equation}
and was shown to reproduce the existing data quite well. We shall  
perform calculations using this form of $G_M^n(Q^2)$ and the 
dipole form of $G_E^n(Q^2)$. In this case however, 
we evaluate the potential numerically and use it to evaluate the
binding energy as explained in the next section.

\subsubsection{The bump tail model}
All new and old data of the electric and magnetic form factors of the
proton and the neutron were recently re-analyzed by a fit 
\cite{walch} based on a pion cloud model. A common feature 
(a bump) was noticed in the data at low momentum transfer, which the authors
attributed to the existence of a pion cloud around a bare nucleon. The 
form factors were parameterized in terms of a bump on top of a large 
smooth part. Purely phenomenologically, the smooth part was parameterized
with a superposition of two dipoles:
\begin{equation}
G_S(Q^2) = {a_{10} \over (1 + Q^2/a_{11})^2} \, +\,
{a_{20} \over (1 + Q^2/a_{21})^2}
\end{equation}
and the bump as a superposition of two Gaussians:
\begin{equation}
G_b(Q^2) = \exp\biggl[-{1\over 2}\,\biggl({Q-Q_b \over \sigma_b}
\biggr)^2 \,\biggr ] \,\,+
\exp\biggl[-{1\over 2}\,\biggl({Q+Q_b \over \sigma_b}
\biggr)^2 \,\biggr ] \, .
\end{equation}
The neutron and proton,  
electric and magnetic form factors are then given by the ansatz,
\begin{equation}
G_N(Q^2) = G_S(Q^2) + a_b \, Q^2\,G_b(Q^2)
\end{equation} 
where the parameter $a_b$ is essentially the amplitude of the bump. 
The parameters for the proton form factors are taken from Table II of 
\cite{walch}
and those for the neutron are taken from a more recent work 
\cite{glazier} where the authors
use a similar phenomenological fit. The neutron parameters fitted in
\cite{glazier} are within the error bars of those quoted in Table II of 
\cite{walch} for the neutron. 

\subsection{ The nuclear form factor}
The form factor of $^{10}$Be is evaluated by taking the Fourier transform of 
a semiphenomenological nuclear charge density given in \cite{gambhir}. 
Though the actual density of $^{10}$Be may have a complicated structure,
the following form provided by the authors in \cite{gambhir} is in 
very good agreement with experimental data on nuclear form factors.
The charge density distribution is given as,
\begin{equation}
\rho (r) = {\rho^0 \over 1 + [(1+(r/R)^2)/2]^\alpha[e^{(r-R)/a} + 
e^{-(r+R)/a}]}\, ,
\end{equation}
where, $\rho^0$ is determined from the normalization condition:
\begin{equation}
4 \pi \,\int\,\,\rho (r) \,r^2\,dr\, = \, Z\, ,
\end{equation}
with $Z$ being the total number of protons in the nucleus. 
The parameters
for $^{10}$Be can be found in Table 1 of \cite{gambhir}.

\section{Corrections to nuclear binding energies}
As mentioned in the previous section, 
we evaluate the correction $\Delta E$ to the binding energy of 
the deuteron and the one-neutron
halo nucleus $^{11}{\rm Be}$. 
\begin{equation}\label{energy}
\Delta E =  \int \, \Psi^*({\bf r}) V^n(r) \Psi ({\bf r}) 
d{\bf r}\, ,
\end{equation}
where $\Psi$ is the wave function corresponding to the unperturbed
Hamiltonian, $H_0$ (the total $H = H_0 + V$).

\subsection{The deuteron}
In the deuteron case, we perform the calculation for the dominant
$s$-state of the wave function obtained using the neutron-proton 
strong interaction. 
When the potential, $V^n(r)$, depends only on the magnitude 
of ${\bf r}$ and is spin-independent, 
the above equation reduces to the simple form:
\begin{equation}\label{energy2}
\Delta E \, = \, \int_0^{\infty} \, r^2 \, u^2(r) \, V^n(r) \, dr 
\end{equation}
where $u(r)$ is the radial part of the deuteron wave function. We use a 
parameterization of this wave function using the Paris nucleon-nucleon 
potential as given in \cite{paris}. Since the Paris potential itself is 
written as a discrete superposition of Yukawa type terms, the wave function 
is parameterized in a similar way as:
\begin{equation}
u(r) = \sum_{j = 1}^{13} C_j\, \exp (-m_j r) \,/\,  r \, .
\end{equation}
The coefficients $C_j$ with dimensions of [fm$^{-1/2}$] are listed in 
Table 1 of \cite{paris} and the masses $m_j$ are given as 
$m_j = \alpha + (j - 1) m_0$, with $m_0 = 1$ fm$^{-1}$ and 
$\alpha = 0.2316$ fm$^{-1}$. In Table I, we list the corrections to the 
deuteron binding energy due to each of the terms in (\ref{fullpotnp}) using the 
most recent parameterization of Refs. \cite{walch} and \cite{glazier}. 
It can be seen that with the $F_1^n F_1^p$ combination of form factors,
all terms other than the leading one are one or two orders of magnitude 
smaller as expected from the ($1/c^2$) expansion. The contributions of
the terms involving $F_2^n(Q^2)$ are large as compared to those
with $F_1^n(Q^2)$ simply due to the fact that unlike $F_1^n(Q^2)$, 
$F_2^n(Q^2)$ does not vanish as $Q^2 \to 0$. 
Table I has been arranged according to the different
form factor combinations to display in a clear way, the theoretical
expectations discussed below Eq. (\ref{fullpotnp}). First, one should 
note that the entry corresponding to the $F_1^n F_1^p$ form factor 
combination has a leading term followed by ($1/c^2$) suppressed next-to-leading
order terms. We have done this to display in this particular case, 
the significance of the next-to-leading order terms. 
In the next three combinations, 
only the leading terms of these particular combinations are listed.
These terms contain a ($1/c^2$) factor 
which, however, does not make them always suppressed as compared
to the leading term of the first combination $F_1^n F_1^p$, 
bearing in mind that we introduce a new
momentum dependent coupling constant $F_2$. This is to say that one
should not expect terms with different form factor combinations to
give similar results.  
This is in agreement with the
remarks we made regarding the $1/c$ expansion in section \ref{subsec:np_case}.
\begin{table}
\caption{\small Corrections due to individual terms in the 
neutron-proton electromagnetic potential, 
to the deuteron binding energy in keV, using 
the parameterizations of form factors in \cite{walch,glazier}.
The corrections assuming a point proton are listed in the last column.}
\vspace{0.5cm}
\label{tab:1}
\renewcommand{\arraystretch}{1.2}
\begin{tabular}{|c|c|c|c|}
\hline
Form factor & 
Term in $U^{np}(\vec{p}_1,\vec{p}_2,\vec{Q})$ &  
{$\Delta E$ (keV)} & $\Delta E$ (keV)(with $F_1^p=1, F_2^p=\kappa_p$)\\[-1.5ex]
combination &  & & \\ \hline
$F_1^n \,F_1^p$ & ${\displaystyle {1 \over \vec{q}^{\,2}}}$ & $-1.25$& $-0.923$   \\ 
    & ${\displaystyle -\,{1 \over 8 \,M_n^2\,c^2}\, -\,
{1 \over 8 \,M_p^2\,c^2}}$     & $-0.016$ & $-0.049$ \\
 &${\displaystyle -\, {\vec{\sigma}_1 \cdot \vec{\sigma}_2
\over 4 \,M_n\,M_p\,c^2}}$& $-0.016$  & $-0.049$\\
 & ${\displaystyle {(\vec{\sigma}_1 \cdot \vec{q})
(\vec{\sigma}_2 \cdot \vec{q}) \over 4 \,M_n\,M_p\,c^2 \vec{q}^{\,2}}} $& $-0.028$
   & $-0.063$\\ 
 &${\displaystyle  -\, {\vec{p}_1\cdot \vec{p}_2 \over M_n\, M_p\,
c^2\,\vec{q}^{\,2}}}$ &  $-0.02$& 0.053 \\ 
 &${\displaystyle {(\vec{p}_1 \cdot \vec{q})\,(\vec{p}_2 \cdot \vec{q})
\over M_n\,M_p\,c^2\,\vec{q}^{\,4}}}$ & $-0.05$& $-0.155$   \\[1.2ex]  \hline
$F_1^n\,F_2^p$ & ${\displaystyle -\,{1 \over 4\,M_p^2\,c^2}}$ & $-0.034$& $-0.177$  \\ 
 &${\displaystyle -\,{\vec{\sigma}_1 \cdot \vec{\sigma}_2
\over 4 \,M_n\,M_p\,c^2}}$ & $-0.034$& $-0.177$   \\
 &${\displaystyle  {(\vec{\sigma}_1 \cdot \vec{q})
(\vec{\sigma}_2 \cdot \vec{q}) \over 4 \,M_n\,M_p\,c^2 \vec{q}^{\,2}}}$ & 
$-0.039$& $-0.1135$ \\ [1.2ex]
 \hline
$F_2^n\,F_1^p$ &${\displaystyle -\,{1 \over 4\,M_n^2\,c^2}}$ &  3.54& 2.986 \\ 
 &${\displaystyle -\,{\vec{\sigma}_1 \cdot \vec{\sigma}_2
\over 4 \,M_n\,M_p\,c^2}}$ & 3.54& 2.986 \\ 
 &${\displaystyle {(\vec{\sigma}_1 \cdot \vec{q})
(\vec{\sigma}_2 \cdot \vec{q}) \over 4 \,M_n\,M_p\,c^2 \vec{q}^{\,2}}}$ & $-1.32$&
 $-2.273$ \\ [1.2ex]
\hline
$F_2^n\,F_2^p$ &${\displaystyle -\, {\vec{\sigma}_1 \cdot \vec{\sigma}_2
\over 4 \,M_n\,M_p\,c^2}}$ & 6.56& 5.355  \\ 
 &${\displaystyle  {(\vec{\sigma}_1 \cdot \vec{q})
(\vec{\sigma}_2 \cdot \vec{q}) \over 4 \,M_n\,M_p\,c^2 \vec{q}^{\,2}}}$ & $-2.02 $
& $-4.076$ \\ [1.2ex]
\hline
 & Total:  & 8.81& 3.3735  \\ \hline
\end{tabular}
\end{table}

In the last column of Table I, we list the corrections assuming
the structured neutron and point proton interaction. The potential
in this case increases a lot in magnitude (for example, the depth of
the potential in the leading term is about $-270$ keV assuming a point
like proton as compared to the $-60$ keV in the
structured proton case), however, the overlap of the
deuteron wave function and the potential does not change drastically.

In order to demonstrate the typical 
form of the neutron-proton electromagnetic potential, 
in Fig. 2, we plot $V^n(r)$ as a function of $r$ for two terms which 
contribute to the binding energy correction with opposite signs.
The solid line corresponds to the attractive potential of the 
$F_1^n(Q^2)\,\,F_1^p(Q^2)$ leading term and the dashed line to the
spin-independent $F_2^n(Q^2)\,\,F_1^p(Q^2)$ term. The other major 
contributors to the total correction $\Delta E$ are spin-dependent and
contain operators. However, if one evaluates the effective potentials  
corresponding to these terms as
mentioned in the appendix, these potentials are seen to have 
a similar form and range as the ones shown in the figure. The effective
potential corresponding to the biggest contribution coming from 
the $F_2^n(Q^2)\,\,F_2^n(Q^2)$ is repulsive with $V^n(r\to0) \sim 220$ keV.  
\begin{figure}[h]
\includegraphics[width=8cm,height=10cm]{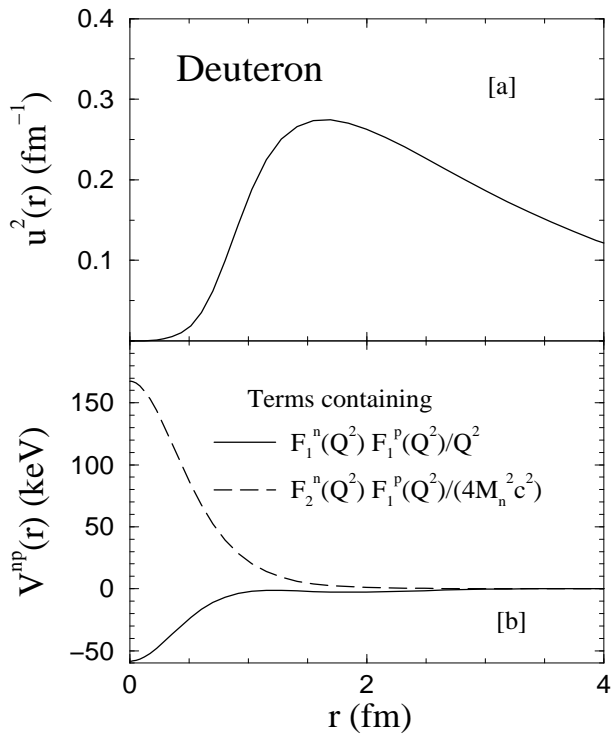}
\caption{\small (a) The $l=0$ radial wave function squared of the deuteron
using the Paris potential and 
(b) two different terms of the electromagnetic potential 
between the neutron and the proton 
evaluated using the parameterization in \cite{walch,glazier}.}
\end{figure}

In Table II, we list the 
total $\Delta E$ due to all terms using the parameterization of Ref. 
\cite{bosted} with different choices of the parameters $a$ and $b$ in 
(\ref{bostedge}).  
The first parameter set in this table, with $a = 0$ corresponds 
to the option $G_E^n (Q^2) = 0$ \cite{footnote}. 
The second set with $a=1$, $b=5.6$ is
the well-known Galster parameterization \cite{gals}. The values in the 
next two rows lie within error bars of the best
fit values obtained in \cite{plat}, namely, $a=1.25 \pm 0.13$ and 
$b = 18.3 \pm 3.4$. The next two rows lie within the error
bars of $a=0.94$ and $b=10.4 \pm 0.6$ which were obtained by
constraining the slope of $G_E^n$ to match the thermal neutron 
data \cite{bosted}. Finally, the last value corresponds to the one
used in \cite{buchman} in connection with the electromagnetic 
$N \to \Delta$ transition.

\begin{table}
\caption{\small Corrections to the deuteron binding energy, $\Delta E$ 
in keV, using the parameterization of form factors as in \cite{bosted}.
$G_{E,M}^p (q^2)$ is taken as in Eqs (4), (5) and 
$G_M^n (q^2)$ as in Eq. (7) in \cite{bosted}. 
The parameters $a$ and $b$ appear in $G_E^n(Q^2)$.}
\vspace{0.5cm}
\label{tab:2}
\begin{tabular}{|c|c|c|}
\hline
 &  &  \\[-2ex]
$a$ &$b$  & $\Delta E $ (keV)  \\ 
 &  &  (Total) \\ \hline
0~~ & - ~~       & ~~7.9 ~~\\ 
1~~ & 5.6 ~~     & ~~11.24~~  \\
1.25~~ & 18.3 ~~ & ~~11.8~~  \\
1.12~~ & 21.7 ~~ & ~~11.39~~  \\ 
0.94~~ & 10.4 ~~ & ~~11.01~~ \\ 
0.94~~ & 11   ~~ & ~~11~~ \\ 
0.9~~ & 1.75  ~~ & ~~10.83~~ \\ [2ex]
\hline
\end{tabular}
\end{table}
It is interesting to note that these corrections to the deuteron 
binding energy are comparable in order of magnitude to those due 
to the presence of a $\Delta$-isobar component in the deuteron. In
\cite{holinde}, the corrections due to the $\Delta$ were found to be 
around 3 keV. 
The deuteron is a precision tool of nuclear physics, both experimentally
(the binding energy is precisely known \cite{greene}) 
and theoretically (as it is a two body problem). 
The level of accuracy demands even to discuss and agree upon
corrections of the order of eV. For example, in \cite{doppler}, 
the so called ``Doppler broadening of the $\gamma$ ray" arising basically
due to the kinetic energy of the neutron in the $^1$H(n,$\gamma$)$^2$H 
reaction (used to extract the deuteron binding energy) 
is found to introduce an error of $25$ eV, maximally. 
This correction is considered by the authors in \cite{doppler}, 
to be a large one as compared to the $2.3$ eV 
error which comes from the $\gamma$-ray detector itself.
The $10$ keV electromagnetic correction is then quite large as 
compared to the above corrections. 
To obtain more accurate estimates of the electromagnetic 
corrections, 
it is therefore clear that there is a need to know the 
neutron form factors to a better precision and for a larger range of the 
momentum transfers than what we know today.

There is a non-zero, albeit small effect of this electromagnetic correction
to the physics of nucleosynthesis. Assuming
that some fundamental constants, 
among them the fine structure constant $\alpha$, change
with the cosmological epoch \cite{changingalp}, 
this variation can affect the abundances
of primordial light nuclei. In \cite{laplata} such an analysis 
for the deuteron gives
\begin{equation} \label{mareknew1}
\frac{\delta Y_d}{Y_d}=2.320\frac{\delta \alpha}{\alpha} + ...
\end{equation} 
where the dots indicate the contributions of the variations of other constants.
The numerical value 2.320 above includes all sources of the $\alpha$ dependence of the
deuteron abundance, except for the direct dependence of 
the binding energy $B_d$ on $\alpha$. To estimate this effect on 
$\delta Y_d/Y_d$
we make use of the equilibrium solution \cite{arnett}
for the deuteron abundance given by
$Y_d \propto \exp[{-B_d/k_BT}]$ and write $B_d=B_d^0 + \alpha B_{em}'=  
B_d^0+ B_{em}$, to
obtain
\begin{equation} \label{mareknew2}
\left(\frac{\delta Y_d}{Y_d}\right)_{\rm direct}=\frac{\delta \alpha}{\alpha}
\,\,\biggl ( -\,\frac{B_{em}}{k_BT} \biggr )\,.
\end{equation}
As an estimate we take, $k_BT \sim 5 \times 10^{-2}$ MeV, corresponding to the 
point where the equilibrium solution deviates from the numerical one due 
to the depletion of the deuteron in the production of heavier nuclei. 
Note however, that the final abundance does not differ much from the
equilibrium solution. This gives $B_d/k_BT  \sim 0.2$ which changes
the coefficient 2.320 in (\ref{mareknew1}) to 2.120. 
The analysis of $\delta Y_d/Y_d$ helps in searching for positive
signals of the variation of fundamental constants.
  
\subsection{One neutron halo - $^{11}{\rm Be}$}
\begin{table}
\caption{\small Corrections to the $^{11}$Be binding energy, $\Delta E$ 
in keV, using the parameterization of neutron form factors as in \cite{bosted}.
$G_M^n (q^2)$ is as in Eq. (7) in \cite{bosted}. 
The parameters $a$ and $b$ appear in $G_E^n(Q^2)$.}
\vspace{0.5cm}
\label{tab:3}
\begin{tabular}{|c|c|c|c|c|}
\hline
 &  & & &\\
$a$ &$b$  & $\Delta E \,$ (keV)& $\Delta E \,$ (keV) 
&$\Delta E \,$ (keV) (Total) \\ 
 &  & due to the & due to the &\\
 &  & term containing & term containing &\\
 &  & $F_1^n(Q^2)$ & $F_2^n(Q^2)$& \\ \hline
0~~ & - ~~       & ~~-3.52 ~~&~~3.53 ~~ &~~0.01~~  \\ 
1~~ & 5.6 ~~     & ~~-0.54 ~~&~~3.55 ~~ &~~3.01 ~~ \\
1.25~~ & 18.3 ~~ & ~~-0.12 ~~&~~3.56 ~~&~~3.44~~ \\
1.12~~ & 21.7 ~~ & ~~-0.54 ~~&~~3.55 ~~&~~3.01 ~~ \\ 
0.94~~ & 10.4 ~~ & ~~-0.82 ~~&~~3.55 ~~&~~2.73 ~~ \\ 
0.94~~ & 11   ~~ & ~~-0.83 ~~&~~3.55  ~~&~~2.72 ~~ \\ 
0.9~~ & 1.75  ~~ & ~~-0.74  ~~&~~3.55 ~~&~~2.81 ~~ \\ 
 &  &  & & \\ \hline
&Ref.\cite{walch,glazier}&~~-1.32~~  &~~3.17~~ &~~1.85~~ \\ 
&Ref.\cite{walch,glazier} ($F_{^{10}{\rm Be}}=4$)&~~$-3.93$~~  &~~17.93~~ &~~14~~ \\ 
 &  &  & & \\ \hline
\end{tabular}
\end{table}
The wave function for $^{11}{\rm Be}$ is taken from a coupled channel 
calculation \cite{filo} performed for one neutron halo nuclei.  
A deformed Woods-Saxon potential for the neutron-core interaction 
is used to take into account the excitation of the 
$^{10}{\rm Be}$ core. The coupling of the neutron to the $^{10}{\rm Be}$ 
core (taken as $0^+$ or $2^+$) 
gives rise to the three components, $2s_{1/2}$, $1d_{3/2}$ and 
$1d_{5/2}$ of the $^{11}{\rm Be}$ wave function. The notation used is 
$nl_j$ with $l$ and $j$ being the orbital and total angular momenta, 
respectively. The normalization of the radial wave functions is such
that $\int [u(r)]^2 dr = 0.85, 0.02$ and $0.13$ for the 
$2s_{1/2}$, $1d_{3/2}$ and $1d_{5/2}$ waves, respectively. The
total wave function for $^{11}{\rm Be}$ as given in 
\cite{filo} is expressed as a sum over spins written 
in terms of the rotational matrices. However, since we have evaluated 
the potential for the spin-1/2 spin 0 case of a neutron and nucleus,
we shall be presenting results only due to the $s$-wave component.
All components of the wave function are however plotted in Fig. 3, 
where one can see that at least for the leading spin-independent term,
the $d$-waves would have a much smaller overlap with the potential as 
compared to the $s$-wave. 

For a detailed comparison of this neutron-$^{10}$Be core model with
the shell model for $^{11}$Be 
as well as experiments, we refer the reader to the 
work of F. M. Nunes {\it et al} \cite{filo}.

In Table III, we list the corrections to the $^{11}$Be binding energy of 
$500$ keV 
using different parameterizations of the nucleon form factors 
as in the case of the deuteron.
We also list the results in the parameterization of \cite{walch,glazier} for
the case assuming a point like charged nucleus.
One can see that the binding energy 
correction turns out to be an order of magnitude
larger than the one which takes the nuclear structure into account.
This big difference (as compared to the not so large one
in the deuteron case with point like proton) can be understood in terms 
of the fact that the 
nuclear form factor falls more rapidly with momentum and the wave function
of the halo nucleus has a different behavior as compared to that of
the deuteron. Thus, if the nuclear form factor is not included, even 
the order
of magnitude of the resulting corrections is incorrect.

Note that in performing the calculation of the
electromagnetic correction for the one-neutron halo, we assumed
the same approach as taken for the strong interaction part, i.e., we
treat the neutron to be interacting with the rest of the nucleus as a whole. 
\begin{figure}[h]
\includegraphics[width=8cm,height=10cm]{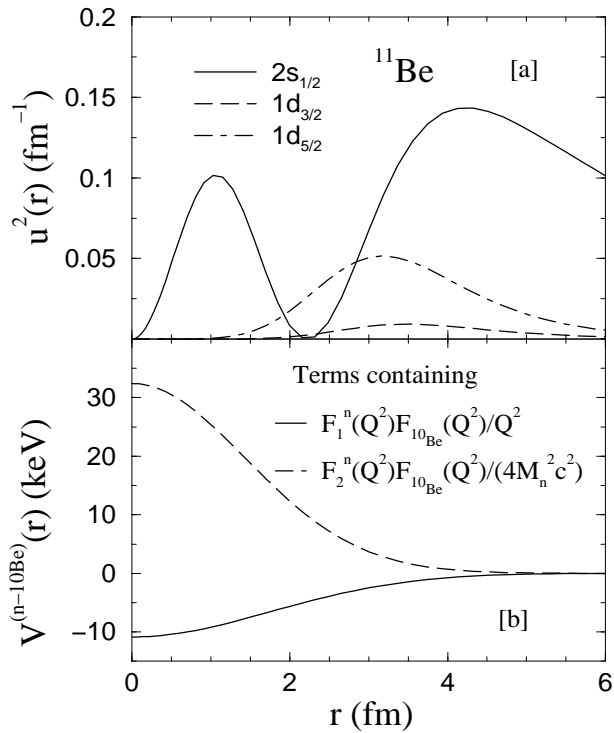}
\caption{\small (a) The radial wave function 
components squared for $^{11}$Be within the neutron-$^{10}$Be-core 
model \cite{filo}.
(b) the electromagnetic potential between the neutron and the 
$^{10}$Be core, 
evaluated using the parameterization in \cite{walch,glazier}.}
\end{figure}

\section{Summary}
To summarize the findings of the present work, we can say that we have
adopted a new approach to the problem of 
calculating electromagnetic corrections to the binding
energies of loosely bound neutron systems encountered in the deuteron 
and one-neutron halo nuclei.
The relevance of such corrections for the deuteron lies in the realm
of precision nuclear physics as $B_d$ is an accurately known number.
A small correction to the analysis of the 
variation of primordial deuteron abundances
with respect to the change of fundamental constants can also be found.
To be able to perform the calculation with a reasonable accuracy we generalized the
Breit equation by allowing the couplings constants to vary 
with momentum transfer (form
factors). To compare the result with other loosely bound systems, 
we calculated the
electromagnetic correction for the one-neutron halo nucleus too. 
This calculation derives its relevance also from
the exotic nature of the halo nuclei. 
Given the low binding energies of the one-neutron halos and the 
fact that in the standard picture, 
the valence neutron resides far from the remaining core,
one could naturally ask the question, does the electromagnetic 
interaction play some role apart from the strong interaction which
is not so strong as in normal nuclei? 
The answer is that the electromagnetic corrections viewed as a ratio to
the corresponding binding energies are of the same order
of magnitude in the case of the deuteron and the one-neutron halo provided
we treat the latter as a neutron interacting with the rest of the nucleus 
(we mentioned already that this is in the same spirit as 
the calculation done for the strong interaction part). 
The strength of the electromagnetic potential between the neutron and the 
core depends on the
number of protons in the core. Hence, such a correction in the case of the 
one-neutron halo $^{19}$C would be larger; more so due to the fact that 
the binding energy of the $^{19}$C can be smaller $\sim 240$ MeV \cite{rid}.

All this also shows that we ought to know the electromagnetic form factors
of the neutron more precisely. 
\vskip0.5cm
{\large \bf Acknowledgment}\\
The authors are grateful to Filomena Nunes for providing the 
$^{11}{\rm Be}$ wave functions required for the calculations of 
the present work. T.M. acknowledges the support from the Faculty of 
Mathematics and Sciences, UI, as well as from the Hibah Pascasarjana.

\newpage
\appendix
\section{The potential in coordinate space}

As mentioned in the main text, the potential in $r$-space is evaluated 
as a Fourier transform (FT) of $U(\vec{p}_1,\vec{p}_2, \vec{Q})$. 
In what follows, we shall enumerate the terms in the order in which
they occur in (\ref{fullpotnp}).
The FT of the first three spin-independent terms 
in (\ref{fullpotnp}) is trivial and gives,
\begin{eqnarray}
V_1(r)&=&{2 \,e^2 \over \pi r} \, \int_0^{\infty} \,\,F_1^n(Q^2)\,\,F_1^p(Q^2)\,
\,{\sin(Qr)\over Q}\,\,dQ \\ 
\nonumber
V_{2(3)}(r)&=&{2 \,e^2 \over \pi r}\,\biggl (-\,{1 \over 8\,M_{n(p)}^2\,c^2} 
\biggr )\, \int_0^{\infty} \,F_1^n(Q^2)\,F_1^p(Q^2)\,Q\,\sin(Qr)\,dQ \,.
\end{eqnarray}
The fourth term which is spin-dependent is evaluated using the relation,
$\vec{S} = (1/2) \, (\vec{\sigma}_1 \,+\,\vec{\sigma}_2)$, leading to
$\vec{\sigma}_1 \cdot \vec{\sigma}_2 = 2\,\vec{S}^2\,-\,3$. Hence,
\begin{equation}
V_4(r) = {2 \,e^2 \over \pi r}\,\biggl ( -{1 \over 4\,M_n\,M_p\,c^2} 
\biggr )\, \int_0^{\infty} \,F_1^n(Q^2)\,F_1^p(Q^2)\,Q\,\sin(Qr)\,dQ \,
(2\,\vec{S}^2\,-\,3)\, .
\end{equation}
The spin operator $\vec{S}^2$ operates on the deuteron spin function 
when one evaluates the binding energy correction as in (\ref{energy}). 
Since the deuteron spin is, $S=1$, $(2\,\vec{S}^2\,-\,3) \,\chi_d = 
+1 \,\chi_d$ and the potential $V_4$ gives a contribution similar to
$V_{2,3}$ up to a factor of about 2. The evaluation of $V_5(r)$ involves
a small mathematical trick as described in \cite{lali4}. The FT of this
term is written as,
\begin{equation}
V_5(r) = \int \, e^{i\vec{Q} \cdot \vec{r}}\, f(Q^2) \, {4 \pi \over Q^2} \, 
(\vec{\sigma}_1 \cdot \vec{Q})\,(\vec{\sigma}_2 \cdot \vec{Q})\, {d^3Q \over 
(2 \pi)^3}\, ,
\end{equation}
where $f(Q^2) = F_1^n(Q^2)\,F_1^p(Q^2)\,e^2/(4\,M_n\,M_p\,c^2)$. 
Now, 
\begin{equation}
V_5(r) = - i \,\sigma_1 \cdot \vec{\nabla}\, \tilde{F}(\vec{r})\,,
\end{equation}
with
\begin{equation}
\tilde{F}(\vec{r}) = \int \, e^{i\vec{Q} \cdot \vec{r}}\, f(Q^2) 
\, {4 \pi \over Q^2} \,(\vec{\sigma}_2 \cdot \vec{Q})\, {d^3Q \over 
(2 \pi)^3}\,.
\end{equation}
Repeating the same trick again, we can write $V_5(r)$ as,
\begin{equation}
V_5(r) = -\,
(\vec{\sigma}_1 \cdot \vec{\nabla})\,(\vec{\sigma}_2 \cdot \vec{\nabla})\, 
\int \, e^{i\vec{Q} \cdot \vec{r}}\, f(Q^2) \, {4 \pi \over Q^2} \, 
{d^3Q \over (2 \pi)^3}\, .
\end{equation}
If we use the relation, 
$2(\vec{S} \cdot \vec{\nabla})^2 \,-\,\vec{\nabla}^2 = 
(\vec{\sigma}_1 \cdot \vec{\nabla})\,(\vec{\sigma}_2 \cdot \vec{\nabla})$, 
and we choose, $\vec{r} = r\,\hat{z}$, then the potential involves 
the operator $S_z^2$, which in the case of the deuteron again acts
on the spin-1 wave function. We can then write an effective potential
(which can be used as in Eq. (\ref{energy2}) for evaluating $\Delta E$) as,
\begin{equation}
V_5^{eff}(r) = {2 \over 3 \pi} \, \int_0^{\infty} \,\,dQ\,\,f(Q^2)\,\,\biggl [\,
\sin(Qr)\,\biggl (-{Q \over r} + {2 \over Q r^3} \,\biggr ) \,-\, 
{2 \over r^2}\,\,\cos(Qr)\,\biggr ]\,.
\end{equation}
The next four terms which involve cross products do not
contribute to the binding energy correction. For example, writing 
the sixth term as, 
\begin{equation}
V_6(r) = - 4 \pi e^2\,i\, \int \, e^{i \vec{Q} \cdot \vec{r}}\, 
{F_1^n(Q^2)\, F_1^p(Q^2) \over 4 \vec{Q}^2 \, M_n^2 \,c^2} \, 
\vec{Q} \cdot (\vec{\sigma}_1 \times \vec{p}_1) \, \, {d^3Q \over (2 \pi)^3}
 \, , 
\end{equation}
and defining,
\begin{equation}
f(\vec{r}) = \int \, e^{i \vec{Q} \cdot \vec{r}}\,
{F_1^n(Q^2)\, F_1^p(Q^2) \over 4 \vec{Q}^2 \, M_n^2 \,c^2} 
{d^3Q \over (2 \pi)^3}\, ,  
\end{equation}
$V_6(r)$ can be expressed as,
\begin{equation}
V_6(r) = - 4 \pi e^2 \,[\,\vec{\nabla} f(\vec{r}) \cdot (\vec{\sigma}_1 
\times \vec{p}_1)\,] \, , 
\end{equation}
where,
\begin{equation}
\vec{\nabla} f(\vec{r}) \cdot (\vec{\sigma}_1 \times \vec{p}_1)\, = \, 
{\partial \over \partial r} f(r) \,\,{x_i \over r} \, \, \epsilon_{ijk} \, \, 
\sigma_{ij}\,\, p_{1k}\, .
\end{equation}
Writing 
$$ p_{1k} \propto {\partial \over \partial x_k} = {\partial \over \partial r}\, 
{\partial r \over \partial x_k} = {x_k \over r}
{\partial \over \partial r } \,, 
$$ 
and the deuteron wave function with its spatial and spin parts as, 
$\Psi = R(r)\, Y_{00}(\theta, \phi)\,\chi$,  
the energy correction as in (\ref{energy}) for $s$-waves becomes,
\begin{equation}
\Delta E = - 4 \pi e^2 \int \, d^3r\,{\partial f(r) \over \partial r} \, \, 
{1 \over r^2}\,\,R^*(r) {\partial \over \partial r}R(r)\,\,
\chi^*\,\sigma_{ij}\,\chi\,\, \epsilon_{ijk}\,x_i\,x_j\,=0\,.
\end{equation}
\noindent
The factor $\vec{p}_1 \cdot \vec{p}_2$ in the next term is given in
the center of mass system of the two nucleons as, $\vec{p}_1 \cdot \vec{p}_2 
\, =\,  -\vec{p}^{\,2} \,=\, \hbar^2 \vec{\nabla}^2$, and acts on the
deuteron wave function in the evaluation of $\Delta E$. Thus,
\begin{equation}
V_{10}(r) = - {2 \, e^2 \over \pi \, r} \, {1 \over M_n \, M_p \, c^2} 
\,\int_0^{\infty} \, F_1^n(Q^2) \, F_1^p(Q^2) \, {\sin (Qr) \over Q} \,dQ\,\,{\partial^2 
\over \partial r^2}\, . 
\end{equation}  
The last of the $F_1^n(Q^2)\,F_1^p(Q^2)$ terms is evaluated using a 
similar mathematical trick as was used for $V_5(r)$. Here again,
working in the center of mass system,
\begin{eqnarray}
V_{11}(r) &=& - (\vec{p}_1 \cdot \vec{\nabla})\,(\vec{p}_2 \cdot \vec{\nabla})
\, \, \int \,e^{i \vec{Q} \cdot \vec{r}}\,\,\biggl [\,{F_1^n(Q^2)\,F_1^p(Q^2)\,
e^2 \over M_n\,M_p\,c^2\,\vec{Q}^2}\,\biggr ]
\,\,{4 \pi \over \vec{Q}^2}\,\, 
{d^3Q \over (2 \pi)^3} \nonumber\\ 
&=& (\vec{p} \cdot \vec{\nabla})^2\, 
\int \,e^{i \vec{Q} \cdot \vec{r}}\,\,\biggl [\,{F_1^n(Q^2)\,F_1^p(Q^2)\,
e^2 \over M_n\,M_p\,c^2\,\vec{Q}^2}\, \biggr ]
\,\,{4 \pi \over \vec{Q}^2}\,\,
{d^3Q \over (2 \pi)^3} \, .
\end{eqnarray}
Here, the momentum operator $\vec{p} = - i \hbar \vec{\nabla}$ acts on
the wave function \cite{lali4} while evaluating $\Delta E$.
The remaining terms in (\ref{fullpotnp}) differ only in the form factor 
combinations and can be evaluated as above.  


\begin{thebibliography}{99}
\bibitem{footnote} The possibility of $G_E^n = 0$ was earlier 
considered as a realistic description of the neutron and we use it here to
display the variation of our results with the different parameterizations.
\end{thebibliography}
\end{document}